\documentclass[sigconf]{acmart}
\AtBeginDocument{%
 \providecommand\BibTeX{{%
 \normalfont B\kern-0.5em{\scshape i\kern-0.25em b}\kern-0.8em\TeX}}}

\setcopyright{acmcopyright}
\copyrightyear{2023}
\acmYear{2023}
\acmDOI{XXXXXXX.XXXXXXX}

\acmConference[SIGIR ECOM '23]{Make sure to enter the correct
  conference title from your rights confirmation email}{July 24--26,
  2023}{Taiwan}
\acmPrice{15.00}
\acmISBN{978-1-4503-XXXX-X/18/06}

\setcopyright{rightsretained}
\acmConference[SIGIR eCom'23]{ACM SIGIR Workshop on eCommerce}{July 27, 2023}{ Taipei, Taiwan}
\acmYear{2023}
\copyrightyear{2023}
\makeatletter
\renewcommand\@formatdoi[1]{\ignorespaces}
\makeatother
\acmISBN{}
\begin{document}

\title{Contextual Font Recommendations based on User Intent}


\author{Sanat Sharma}
\email{sanatsha@adobe.com}

\affiliation{%
 \institution{Adobe Inc.}
 \country{USA}
}

\author{Jayant Kumar}
\email{jaykumar@adobe.com}

\affiliation{%
 \institution{Adobe Inc.}
 \country{USA}
}

\author{Jing Zheng}
\email{jinzheng@adobe.com}
\affiliation{%
 \institution{Adobe Inc.}
 \country{USA}
}

\author{Tracy Holloway King}
\email{tking@adobe.com}
\affiliation{%
 \institution{Adobe Inc.}
 \country{USA}
}


\begin{abstract}

Adobe Fonts has a rich library of over 20,000 unique fonts that Adobe users utilize for creating graphics, posters, composites etc. Due to the nature of the large library, knowing what font to select can be a daunting task that requires a lot of experience. 
For most users in Adobe products, especially casual users of Adobe Express, this often means choosing the default font instead of utilizing the rich and diverse fonts available. 
In this work, we create an intent-driven system to provide contextual font recommendations to users to aid in their creative journey. Our system takes in multilingual text input and recommends suitable fonts based on the user's intent. Based on user entitlements, the mix of free and paid fonts is adjusted. 
The feature is currently used by millions of Adobe Express (shown in Fig.\ \ref{fig:fontreq}) users with a CTR of $>$25\%.
%The feature is currently used by millions of Adobe Express (shown in Fig.\ \ref{fig:fontreq}) users with a CTR of 29\%.

\end{abstract}





\maketitle

\begin{figure}[htb]
 \centering
 
 \includegraphics[width=3.3in]{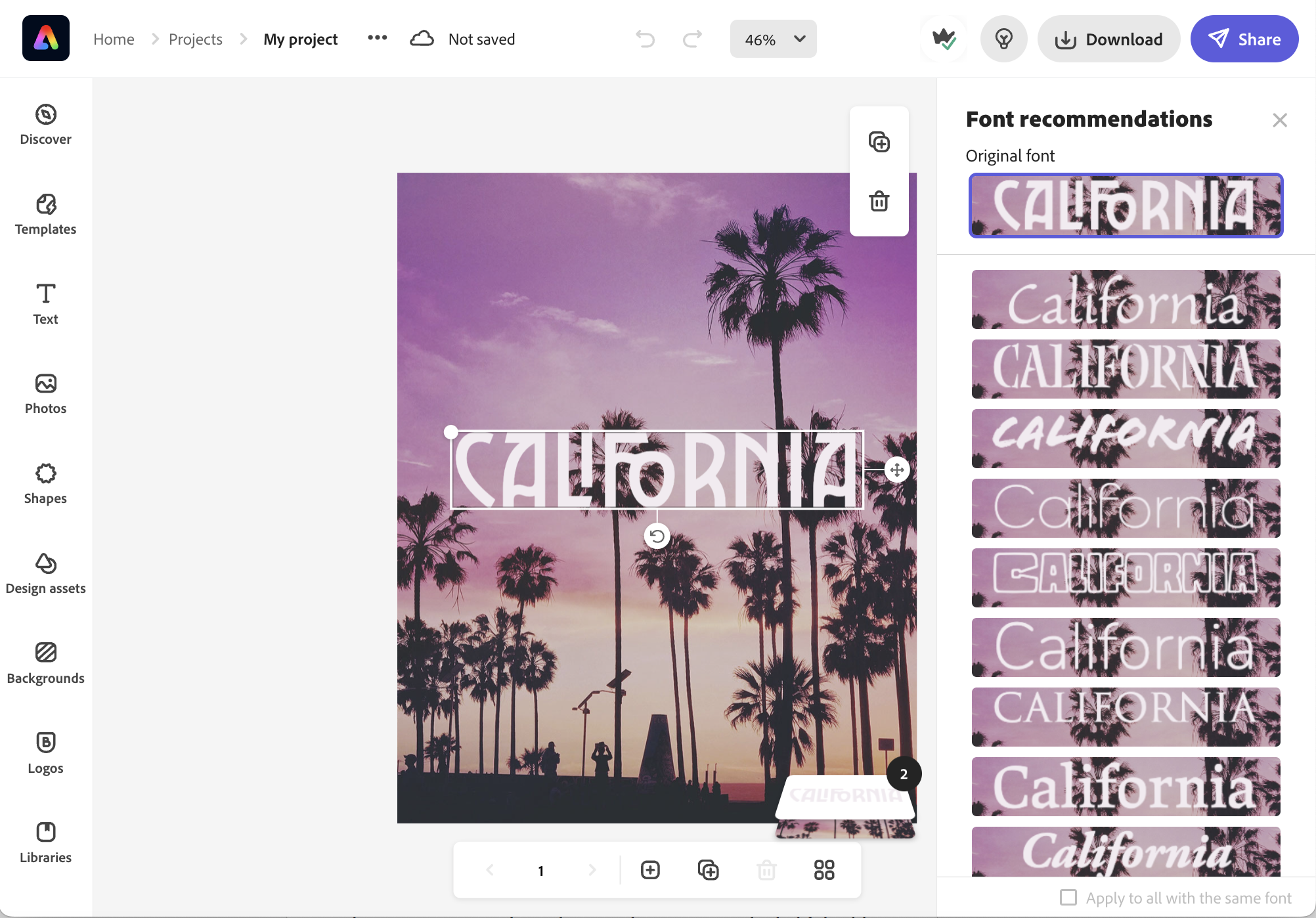} 
\caption{Example font recommendations in Adobe Express.
}
\label{fig:fontreq}
\end{figure}

\section{Introduction: Application}

Fonts, including the difficulty of font search and recommendations, have been a well-studied area in industry and at Adobe in particular.\footnote{We thank Andrei Stefan, Ravindra Sadaphule, and Arpita Agrawal for their invaluable help in productionalizing the font recommendation feature shown in Fig.\ \ref{fig:fontreq}.}\ Previous works have tagged fonts with a small array of tags which can then be used for search. For example, Adobe Fonts ascribes 26 tags to fonts on their site. Recent works have used the MyFonts dataset to assign tags to fonts and utilized them for retrieval \cite{https://doi.org/10.48550/arxiv.1909.02072} \cite{Baird1994Selfcorrecting1C}. There are also projects that do font pairing and retrieval based on visual appearance of the font \cite{font_pair} \cite{DBLP:journals/corr/abs-2103-10036}.

However, doing font retrieval based on font tags often misses the semantics governing what types of contexts the font should be used in. Furthermore, retrieving and recommending fonts based on the visual style of the fonts already on the canvas often leads to poor results when there is variety in the content of the page.

Our work focuses on understanding the intent of the the user's project using multimodal input (text and image context). We then recommend fonts that work well with the project intent. This allows us to understand the semantic relationship between the user's input (e.g.\ the text  they need to select a font for) and the fonts in our library. Our work comprises:

\begin{enumerate}
 \item A multimodal Input $\rightarrow$ Font model that can accurately and reliably recommend a diverse array of fonts to the user, by extracting their creative intent.
 \item A method to recommend ``similar'' fonts by intent, rather than simple appearance.
 \item A scalable query architecture for low latency recommendations.
\end{enumerate}

Since we map the user's context to intent and then to fonts, we are able to showcase a diverse set of fonts, including specialized, unique fonts. This intent-based method also allows us to train our models on a much smaller set of data, which is often a problem in training font recommendation models. The approach is currently used in production in Adobe Express, with a CTR of $>$25\% and a downstream project export rate of $>$50\% whenever a recommended font has been clicked: that is, users who interact with the recommended fonts are very likely to finish and export the project they are working on. Whether the user is using a free, trial, paid, or enterprise account has an effect on engagement with font selection in general and  with subsequent use of the font recommendations. %The approach is currently used in production in Adobe Express, with a CTR of 29\% and a downstream project export rate of 57\% whenever a recommended font has been clicked: that is, users who interact with the recommended fonts are very likely to finish and export the project they are working on.

\section{Data}

In order to mine data for training, we extracted text-font pairs for $\sim$212,000 texts in Adobe Express templates. Adobe Express is a platform for people to create flyers, posters, cards, logos, etc.\ and hence has templates with a diverse set of fonts. Previous font recommendation approaches often utilized general text documents, which rarely have much diversity of fonts \cite{DBLP:journals/corr/abs-2104-10741} \cite{font_pair}. 

For each text input, we determine the top intents via our intent detection model. The intent detection model is trained on 1500+ intents, derived from Adobe Express template tags. These tags range from events such as ``Christmas'' and ``Halloween'' to objects and emotional characteristics such as ``balloon'', ``happy'', and ``encouraging''. This results in triplets of font + text + intent. This, to our knowledge, is the largest dataset of font-text-intent mapping. Fig.\ \ref{fig:font_dataset} showcases a sample of the training dataset. After further thresholding and preprocessing, we extract 470,000 rows of text + font + intent triplets that we utilize to train our models.

\begin{figure}[htb]
 \centering
\begin{tabular}{lllll}\hline\hline
Id & Text & Font & Intent & Score \\\hline
150 & photography & Gizmo & photography & 0.0249 \\
& & & camera & 0.0063\\
& & & photographer & 0.0038\\
& & & woman & 0.0020\\
& & & sunset & 0.0018\\
%693 & SHOW: & Myriad Hebrew & streaming & 0.0005\\
693 & Manufactory & Myriad Hebrew & food & 0.0021\\
& & & business & 0.0013\\
& & & restaurant & 0.0013\\
& & & cafe & 0.0007\\
\hline\hline
\end{tabular}
 \caption{Training dataset of texts, intents and font used}
 \label{fig:font_dataset}
\end{figure}

\begin{figure}[htb]
 \centering

 \begin{tabular}{l}
 \hline\hline
 {\bf Shlop}: fun, spooky, halloween, music, art, scary, party \\
 \includegraphics[width=2.5in]{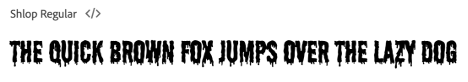}\\
 \hline
 {\bf Rig Solid}: typography, business, fun, education, fashion,\\
 \hspace*{4.5em}music, travel \\
 \includegraphics[width=2.5in]{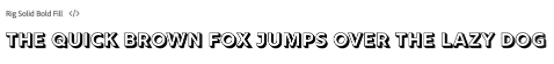}\\
 \hline
 {\bf Sudestada}: wedding, floral, marriage, mom, happy \\ 
 \includegraphics[width=2.5in]{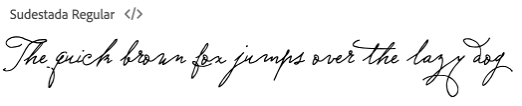}\\
 \hline\hline
 \end{tabular}
 \caption{Font characteristics based on intent: The font name is followed by the related intents (from most to least relevant). An example of the font is shown below the font name.}
 \label{fig:font_charact}
\end{figure}

Since  fonts within a font family are visually similar and have other similar characteristics, we group similarly used fonts in a font family together based on intent. This grouping helps to capture artistic factors of the fonts and corresponding intents. The final training set consists of 2043 unique font families that we recommend from.   Fig.\ \ref{fig:font_charact} shows some examples of fonts and their associated intents.

\section{Model}

\subsection{Intent Detection Model}

\subsubsection{Creative Intent Taxonomy}

In order to accurately map user creative intent, we  semi-automatically extract  a creative intent taxonomy from the Adobe Express content metadata and users' behavioral data when using and searching for content. We utilize Adobe Express templates for this taxonomy, since the template data contains rich designer-added tags and represents a diverse set of creative intents. The intents comprise two main types (example broad and narrower intents in parentheses): 

\begin{enumerate}
\item	Topic of work/task/emotion (e.g.\ yoga, Halloween, cosmetic, food, book launch, happiness, birthday, birthday party)
\item	Type of creation (e.g.\ flyer, poster, social media post, card, greeting card, christmas card, valentine's day card)
\end{enumerate}

We utilize a pretrained DistilBert \cite{https://doi.org/10.48550/arxiv.1910.01108} transformer to cluster related topics and then apply popularity metrics to derive a creative intent taxonomy of  over 1500 intents mined from the Adobe Express dataset. This clustering allows us to collapse terms such as "mlk jr day", "martin luther jr day" and "mlk day" into one concept. We keep the concept with the most popularity (most frequent occurrence among all templates).%, a sample of which can be seen in Fig \ref{fig:taxonomy}.
%\begin{figure}[htb]
 %\centering
 %\includegraphics[width=2.5in]{taxonomy.png}
 %\caption{Creative Intent Taxonomy mined for understanding user behavior.}
 %\label{fig:taxonomy}
%\end{figure}

\begin{figure*}
 \centering
 \begin{tabular}{ll}
 \hline\hline
 \multicolumn{2}{l}{{\bf Anchor Text}: Be excited for spooky Sunday at my apartment.}\\
 {\bf Intent}: spooky & \\
 {\bf Positive}: Brim Narrow Combined & {\bf Negative}: Bio Sans\\
 \includegraphics[width=3in]{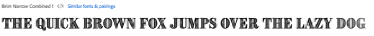} & \includegraphics[width=3in]{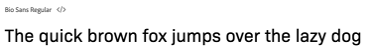}\\\hline
 \multicolumn{2}{l}{{\bf Anchor Text}: We will host a seminar on career development in Welch 102.}\\
 {\bf Intent}: informational & \\
 {\bf Positive}: Journal & {\bf Negative}: Walnut\\
 \includegraphics[width=3in]{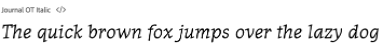} & \includegraphics[width=3in]{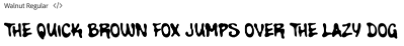}\\
 \hline\hline
 \end{tabular}
 \caption{Triplet pairs used for training. We use intents and positive, negative fonts for training}
 \label{fig:triplet_pair}
\end{figure*}

\subsubsection{Multilingual Text-Intent Model}

Users often add text to their creation to convey  their intent. For example, ``Happy Mother’s Day'' is written on Mother’s Day greeting cards, and ``We stand in solidarity with black businesses'' for BLM social media posts. It is crucial to infer the intent (topic + type) of a user project based on the text entered on the canvas or issued as a search. We trained a transformer model which takes as input a text sample (sentence or phrase or word) and predicts the top, most probable intents from our curated taxonomy. We extracted over 335,000 pairs of text from Adobe Express templates and mapped them to the ``topics'' that were manually tagged by the template designers. Our final dataset consisted of over 1 million text–intent pairs that we utilized to train our classification model. We fine-tuned a DistilBert \cite{https://doi.org/10.48550/arxiv.1910.01108} using contrastive learning to learn the text-to-intent relationships.
Our model works for 36+ languages including English, French, German and Japanese, which is crucial for  any production feature at Adobe.

%% moved fig with triplets up so that it would float better

\subsection{Font Recommendation Model}
In order to model intent and fonts in the same embedding space efficiently, we utilized Triplet Learning and online triplet mining \cite{https://doi.org/10.48550/arxiv.1412.6622}. This effectively generates intent-to-font pairings, as shown in Fig \ref{fig:triplet_pair}.
We fine-tuned a pretrained SimCSE model trained on query understanding \cite{https://doi.org/10.48550/arxiv.2104.08821} and trained the model with Triplet Margin Loss \cite{BMVC2016_119} with a margin of 2. The loss function is summarized in equation \ref{eq:loss} where: $\alpha$ is the margin; $A, P$ and $N$ denote the anchor, positive and negative sample respectively; $f$ denotes the neural network. We utilize Adam \cite{https://doi.org/10.48550/arxiv.1412.6980} as our optimization function with a high learning rate (empirically we found $3e^{-3}$ to give the best results).

\begin{equation}
 L(A, P, N) = max(\parallel f(A)- f(P)\parallel^2 - \parallel f(A) - f(N)\parallel^2 + \alpha, 0)
\label{eq:loss}
\end{equation}

One challenge  was to prevent commonly used fonts from skewing recommendation variation. For generating the embedding for each font, we found the best results by representing a font by a ranked set of its top 7 intents. The ranking is based on the prevalence of the intent-font pair in our dataset. This  helped distinguish unique fonts and  remove font bias at runtime.

\section{Service Architecture}

The system design of font recommendation has three main components: the font recommendation service, Search Platform as a Service (SPaaS) and Universal Search Service (USS). 

The USS service defines the input-output schema for clients and acts as the orchestrator between the font recommendation service and the Search Platform (SPaaS).

SPaaS  contains the metadata, embeddings and intent information across various Adobe assets such as backgrounds, images, fonts and Adobe Express templates. SPaaS is responsible for providing the font recommendation service with  input information (context, embeddings, etc.) present in its Elasticsearch (ES) index as well as extracting font metadata from the ES index for each recommended font returned from the service.

\begin{figure*}[htb]
 \centering
 \includegraphics[width=4in]{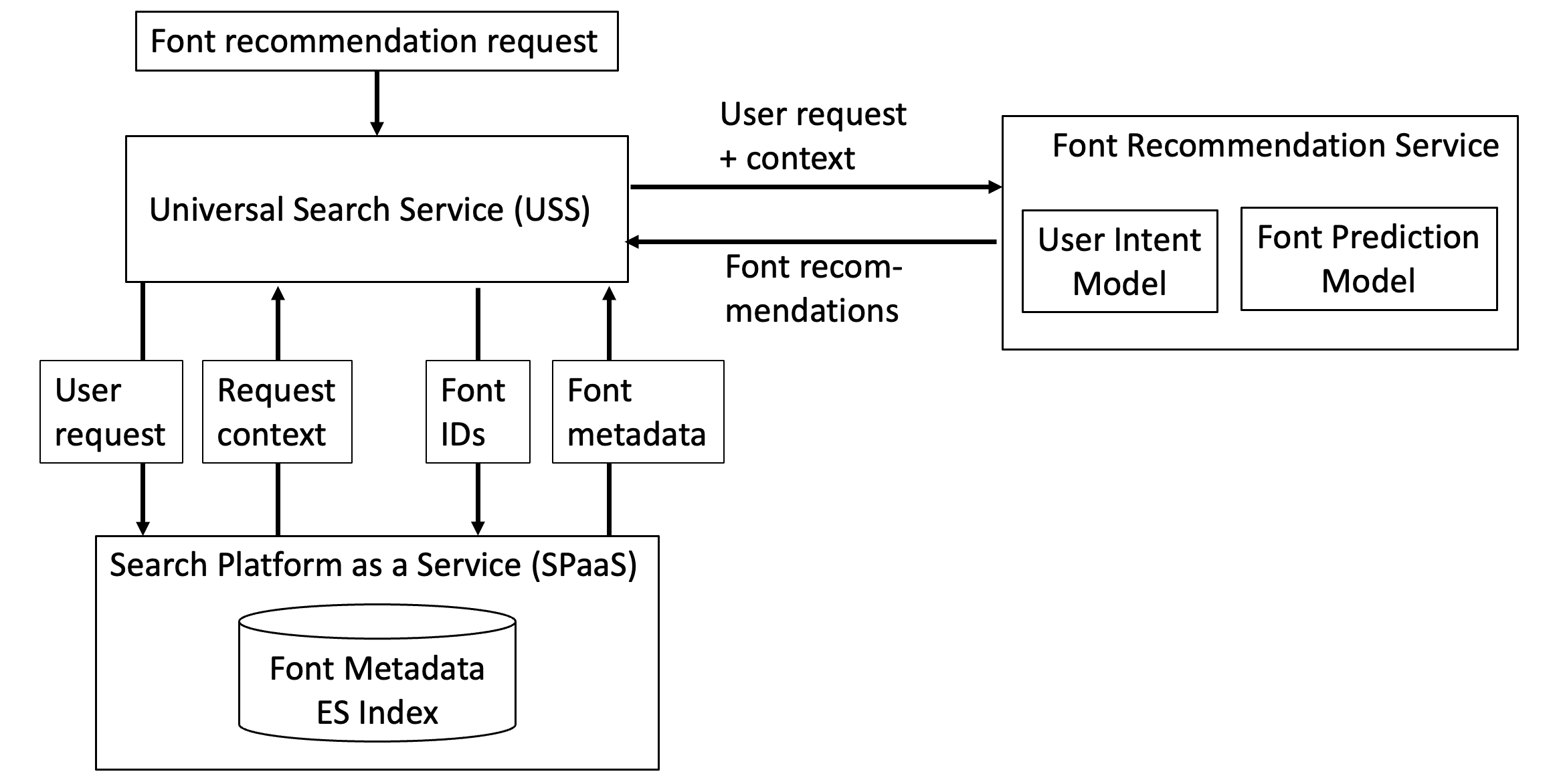}
 \caption{Service Architecture for Font Recommendation}
 \label{fig:arch}
\end{figure*}

At runtime, our clients send user requests (i.e.\ text input from the Adobe Express editing canvas) to USS. USS retrieves additional  information from SPaaS before sending the full payload to the font recommendation service. The font recommendation service extracts the intents from the text input and then uses the weighted set of intents to generate an embedding. This embedding  is  used to select font recommendations. When the input language has a select set of compatible fonts (e.g.\ Japanese), SPaaS provides the recommendation service with the subset of fonts to recommend from. The full architecture is shown in Figure \ref{fig:arch}.

\section{Evaluation}
\subsection{External Manual Evaluation}

%\begin{table}[bht]
%\centering
%\begin{tabular}{|l|l|l|}
%\hline
%Very Good & Ok & Not Good \\ 
%\hline
%464 & 216 & 159
%\\
%\hline
%\end{tabular}
%\caption{Relevancy external evaluation: 75 annotators}
%\label{tab:eval}
%\end{table}

We conducted an external evaluation with 75 evaluators. We selected 78 high frequency templates and displayed them to annotators with one of the texts removed. We then showed the annotators 5 recommended fonts for the text that had been removed and asked then to rate each of the fonts as ``very good'', ``ok'' or ``not good''. Since font recommendations are highly subjective, we asked the annotators to fully consider each font's compatibility with the text. We found that evaluators considered a text-font pair relevant (very good or ok) \textbf{81\%} of the time (Table \ref{tab:eval}).

We also evaluated a weak baseline of randomly selected fonts and a strong baseline of common fonts. Both were $>$10ppt (see fn.\ \ref{fn:ppt}) worse than the intent-based  algorithm. Randomly selected fonts often showed inappropriate, heavily stylized fonts. Popular fonts showed visually similar fonts with little diversity or newness: This reduced discoverability of unique fonts (e.g.\ for  Halloween, where stylized fonts work well, only non-stylized options were shown).

\subsection{Internal Manual Evaluation}

\begin{figure}[htb]
 \centering
 \begin{tabular}{l}
 {\bf Text}: pizza pasta burgers\\
 {\bf Extracted Intents:} Food, takeaway, fast food, restaurant, burger\\
\includegraphics[width=\linewidth]{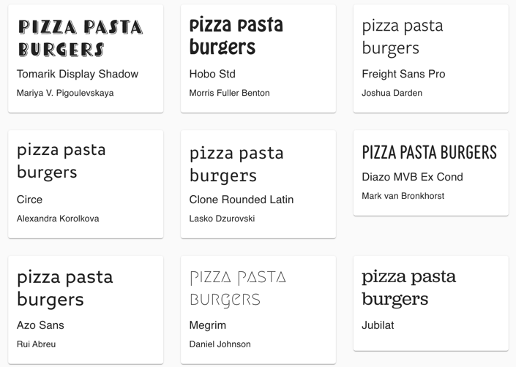}\\
 \end{tabular}
 \caption{Top Text-to-Font recommendations for the text {\bf pizza pasta burgers}. Classic menu fonts for casual restaurants are recommended.}
 \label{fig:exampl}
\end{figure}

We also conducted an internal  evaluation  with 16 annotators before  AB testing the font recommendations. Similar to the external evaluation, we took 100 texts in multiple languages (English, French, German, Japanese) from Adobe templates and asked team members to rank the font recommendations on a scale of 1--5  (1 = very bad, 5 = very good). Our internal results reflected the external evaluation, with a mean score of \textbf{3.67}. Since fonts are so subjective, this was considered sufficient for AB testing. The score distribution is shown in Table \ref{tab:eval}. Some examples of the internal evaluation queries are shown in Figure \ref{fig:exampl}.

%\begin{table}[htb]
%\centering
%\begin{tabular}{|l|l|l|l|l|}
%\hline
%Very Good & Good & Ok & Bad & Very Bad \\ 
%\hline
%10 & 65 & 56 & 11 & 1
%\\
%\hline
%\end{tabular}
%\caption{Relevancy internal evaluation results: 16 annotators}
%\label{tab:eval_internal}
%\end{table}

\begin{table}[htb]
\centering
\begin{tabular}{|l|l|l|}
\multicolumn{3}{c}{External evaluation: 75 annotators}\\
\hline
Very Good & Ok & Not Good \\ 
\hline
464 & 216 & 159\\
\hline
\end{tabular}

\begin{tabular}{|l|l|l|l|l|}
\multicolumn{5}{c}{Internal evaluation: 16 annotators}\\
\hline
Very Good & Good & Ok & Bad & Very Bad \\ 
\hline
10 & 65 & 56 & 11 & 1\\
\hline
\end{tabular}
\caption{Relevancy results for internal \& external evaluations}
\label{tab:eval}
\end{table}

\subsection{In-Production Analysis}
Once the external and internal evaluations were completed, the font recommendations were put in production\footnote{We cannot report exact numbers for proprietary reasons. Instead we provide lower bounds and relative numbers.}\ in Adobe Express in the experience shown in Fig.\ \ref{fig:fontreq}. Overall, the recommendations have a click-through rate (CTR) of $>$25\% and users who select a recommended font have a downstream project export rate of $>$50\%. 

There are 3 types of user accounts on Adobe Express: Free, trial, and enterprise. Although the font recommendation algorithm is identical for all users, the percentage of free and paid fonts depends on their account type. We analyzed the engagement and success of the different  types, focusing on users who edited text since font recommendations are only relevant to such users. Surprisingly, free and enterprise users who edited text had the same rates for using the font selection module, selecting a recommendation, and exporting a final project. In contrast, trial users had much higher engagement at each step (font module +15ppt; recommendation selection +13ppt; export +2ppt).\footnote{We are unable to publish the exact rates. We report the results  as percentage point (ppt) change. For example, a +5 in CTR indicates going from a CTR of X\% (e.g.\ 60\%) to a CTR of X+5\% (e.g.\ 65\%). \label{fn:ppt}}\ 

Users who utilize font recommendations have a higher  project export (+25ppt) compared to users who do not. This may be due to the fact that engaged users have higher desire to complete their projects. This is also consistent with overall project export rates after click on font recommendation for paid (trial + enterprise + non-enterprise paid) vs.\ free users. Paid users had a +10ppt export rate compared to free users after clicking on font recommendations.

We also saw much higher (3x) engagement on  web  compared to mobile surfaces. We attribute this to the fact that the feature is more prominent on web as well as having a larger screen allows users to browse recommendations more easily. 

Finally we  draw some key insights about paid vs.\ free fonts. For paid users, where there is an equal split of free and paid fonts, users click on paid fonts slightly more (+2ppt) than free ones. In contrast, free users are shown a much higher percentage of free fonts and rarely select paid fonts due to the payment barrier. This reflects the fact that while paid fonts are generally better liked, they are not enough incentive  for most users to convert to paying customers.

\section{Conclusion and Next Steps}

This work showcased a novel end-to-end framework for recommending fonts based on multilingual user input. The recommendation system includes a model to understand user intent and brings intents and fonts into the same representation space. It  includes  a low-latency,  scalable architecture to serve the font recommendations. The current approach has high accuracy and relevance in their recommendations and is highly used by Adobe Express users. 

Next steps focus on whole document relevance. Firstly, we want to understand the  fonts already in the document so that we can recommend fonts that work in harmony with the existing artistic design. Secondly, the model described here focuses on individual texts, whereas global intent is not taken into consideration.  Finally, we currently do not support right-to-left languages like Arabic.

\bibliographystyle{ACM-Reference-Format}
\bibliography{SIRIPrefs}

\end{document}